\documentclass[twocolumn]{webofc}

\usepackage[varg]{txfonts}   
\usepackage{graphicx,tikz}

\usepackage[T1]{fontenc}
\usepackage[utf8]{inputenc}

\usepackage{float}

\begin{document}
\title{Fission modes in fermium isotopes with Brownian shape-motion model}

\author{\lastname{M. Albertsson}\inst{1}\fnsep\thanks{\email{martin.albertsson@matfys.lth.se}}  
\and
        \lastname{B.G. Carlsson}\inst{1}
\and
        \lastname{T. D{\o}ssing}\inst{2}
\and
        \lastname{P. M\"oller}\inst{1,3}
\and
        \lastname{J. Randrup}\inst{4}
\and
        \lastname{S. {\AA}berg}\inst{1}
}

\institute{Mathematical Physics, Lund University, 221 00 Lund, Sweden
\and
           Niels Bohr Institute, DK-2100 Copenhagen {\O}, Denmark
\and
           P. Moller Scientific Computing and Graphics, Inc. P.O. Box 1440, Los Alamos, New Mexico 87544, USA
\and
           Nuclear Science Division, Lawrence Berkeley National Laboratory, Berkeley, California 94720, USA
}

\abstract{
Fission-fragment mass and total-kinetic-energy distributions following fission of the fermium isotopes $^{256,258,260}\text{Fm}$
at low excitation energy have been calculated using the Brownian shape-motion model. 
A transition from asymmetric fission in $^{256}\text{Fm}$ to symmetric fission in $^{258}\text{Fm}$ is obtained.
The total-kinetic-energy distributions for the three isotopes show radically different behaviour due to varying contributions from different fission modes,
with a double-humped distribution for $^{258}\text{Fm}$.
}

\maketitle

\section{Introduction}
\label{sec-1}
The co-existence of two modes in fission leading to very different values of the total kinetic energy (TKE) of the fragments, known as bimodal fission, 
has been observed in the fermium region~\cite{hulet1986}.
This phenomenon is argued to arise due to two fission paths in the potential-energy landscape;
one path is described by the liquid-drop model leading to elongated scission configurations with low TKE,
while the other path is influenced by both fragments being close to the doubly magic nucleus $^{132}\text{Sn}$
leading to compact scission with high TKE.

In our endeavour to understand fission modes of nuclei in the superheavy region~\cite{albertsson2019},
we first consider fission in neutron-rich fermium isotopes, 
which is often used as a benchmark for theoretical studies.
We calculate fission-fragment mass and TKE distributions for the
fermium isotopes $^{256,258,260}\text{Fm}$ using the Brownian shape-motion (BSM) model~\cite{randrup2011}.
The calculational details are explained in Sect.~\ref{sec-2}. 
In Sect.~\ref{sec-3} the calculated fission-fragment mass and TKE distributions are presented with a short discussion about the two fission modes.

\section{Calculational details}
\label{sec-2}
Fission-fragment mass distributions are calculated using the BSM method on the five-dimensional
potential-energy surfaces as specified in Refs.~\cite{randrup2011,randrup2013}.
The potential energy of the fissioning nucleus is calculated within the macroscopic-microscopic model~\cite{moller2009}
in a grid of more than 6 million shapes, where the macroscopic part is calculated within the finite-range liquid-drop model
and the microscopic part is calculated with the folded-Yukawa single-particle model.
In the Strutinsky shell-correction procedure we use a larger smoothing range~\cite{albertsson2019} compared to Refs.~\cite{randrup2011,randrup2013}, 
since this is particularly important for nuclei in the vicinity of fermium~\cite{moller1989,moller1994}.
Since most of the CPU time is used to calculate the single-particle levels, we use the same set of levels of $^{258}$Fm to calculate the shell
corrections also for $^{256}$Fm and $^{260}$Fm.

The starting point for the dynamical simulations is in the isomer minimum at low compound excitation energies, just above the outer barrier.
This corresponds to excitation energies $E^\ast=$ 3.88, 2.77 and 1.71 MeV for $^{256-260}\text{Fm}$, respectively.
The shape evolution of the fissioning nucleus is performed using the effective level density of Ref.~\cite{randrup2013} with a bias potential of 60 MeV
in order to decrease the computational time\footnote{Using a bias potential of 30 MeV slightly increases the full width at half maximum in the mass yield for $^{258}$Fm 
by about 8 nucleons while the computational time increases by roughly a factor of 20.}.
The scission point is defined as when the neck radius reaches the critical value $c_0=1.5$ fm.
If the critical neck radius is not reached within $2\times 10^6$ steps or if the walk reaches the maximum
value of the elongation coordinate, then the walk is discarded.
A total of 10000 walks are performed for each nucleus, with less than 1\% of the walks being discarded.

The calculation of the TKE is done as specified in Ref.~\cite{albertsson2019}.
The total available energy in the fission process, given by the sum of the compound nucleus excitation energy $E^\ast$ and the $Q$-value,
is shared between the TKE and the total excitation energy (TXE) of the two fragments.
The TXE is composed of two parts;
intrinsic excitation energy in the compound nuclues at scission and distortion energy 
released in the fragments when they relax their respective scission shapes (taken as the cap deformations in the 5D shape parametrization) 
to their ground-state forms. 

The shape-dependent masses entering in the calculation of the $Q$-values, as well as distortion energies
are calculated in the same macroscopic-microscopic model that was used to obtain the potential-energy surfaces~\cite{moller2004}.
We only consider $\varepsilon_{2},\varepsilon_{4},\varepsilon_{6}$ for the ground-state deformations.

The number of protons $Z$ and neutrons $N$ in a fragment are determined by requiring the same $Z/N$ ratio as for the fissioning nucleus. 
In the present study only fragments with even $Z$ and $N$ are considered. 

\section{Results and discussion}
\label{sec-3}
\begin{figure}[hbt!]
\centering
\includegraphics[width=0.8\linewidth]{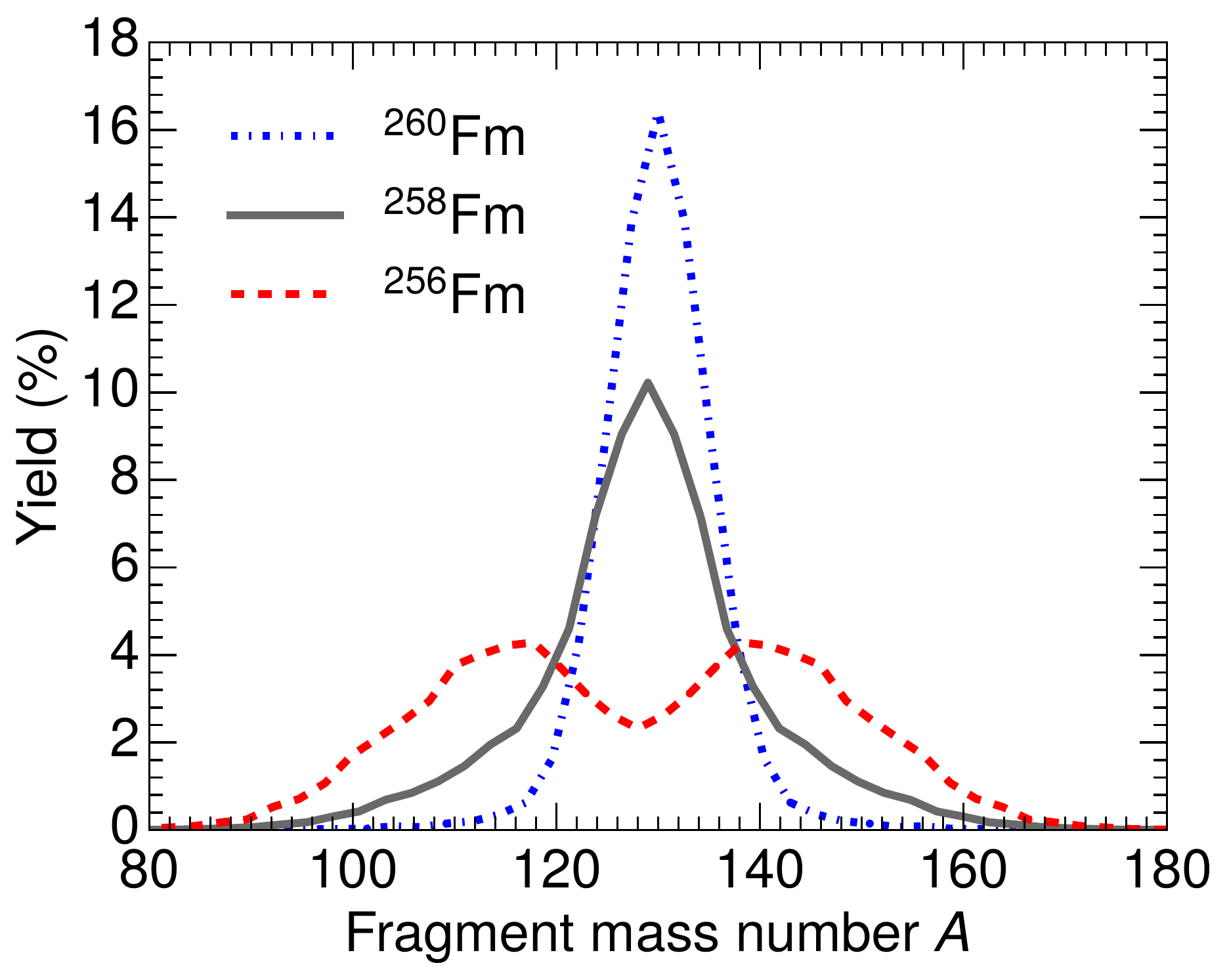}
\caption{Calculated fission-fragment mass yields for the fermium isotopes $^{256,258,260}\text{Fm}$.}
\label{fig-1} 
\end{figure}

The calculated fission-fragment mass distributions for $^{256,258,260}\text{Fm}$ are shown in Fig.~\ref{fig-1}.
A transition from asymmetric fission in $^{256}\text{Fm}$ to symmetric fission in $^{258}\text{Fm}$ is obtained.
This transition is in agreement with experimental data for spontaneous fission~\cite{hulet1986,flynn1972},
and has been ascribed to the strong shell effects emerging from fragments approaching the doubly magic nucleus $^{132}$Sn.

\vspace{-2.8mm}

\vspace{-5mm}

\begin{figure}[hbt!]
\begin{tikzpicture}[remember picture]

\fontfamily{phv}{\fontsize{9}{9}\selectfont

\draw (0,0.5) -- (8,0.5) -- (8,3.5) -- (0,3.5) -- (0,0.5);
\draw (5.33,0.5) -- (5.33,3.5);
\draw (2.67,0.5) -- (2.67,3.5);
\draw (0,2.9) -- (8,2.9);
\filldraw[draw=black,fill=red] (0,2.9) rectangle (2.67,3.5);
\filldraw[draw=black,fill=black!40!white] (2.67,2.9) rectangle (5.33,3.5);
\filldraw[draw=black,fill=blue!50!white] (5.33,2.9) rectangle (8,3.5);
\node[text width=3cm] at (2.35,3.2) {$^{256}$Fm};
\node[text width=3cm] at (5.0,3.2) {$^{258}$Fm};
\node[text width=3cm] at (7.65,3.2) {$^{260}$Fm};
\node[text width=3cm] at (1.95,1.04) {Asymmetric};
\node[text width=3cm] at (2.1,0.7) {Elongated};
\node[text width=3cm] at (4.95,0.85) {Bimodal};
\node[text width=3cm] at (7.40,1.04) {Symmetric};
\node[text width=3cm] at (7.55,0.7) {Compact};
\node[inner sep=0pt] (256fm) at (1.35,2.45) {\includegraphics[width=.12\textwidth]{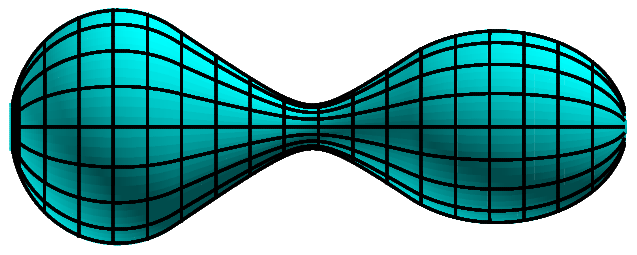}};
\node[inner sep=0pt] (258fm1) at (4.0,2.45) {\includegraphics[width=.12\textwidth]{258fm_elong.png}};
\node[inner sep=0pt] (258fm2) at (4.0,1.65) {\includegraphics[width=.10\textwidth]{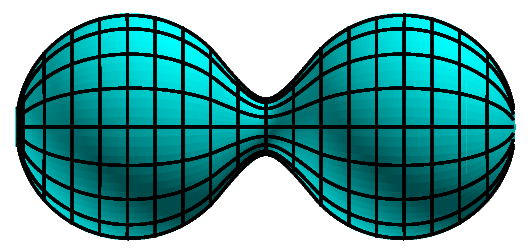}};
\node[inner sep=0pt] (260fm) at (6.65,1.65) {\includegraphics[width=.10\textwidth]{258fm_compact.png}};

}

\end{tikzpicture}
\caption{Typical scission shapes for $^{256,258,260}\text{Fm}$ obtained in the calculations.
}
\label{fig-2}
\end{figure}

Figure~\ref{fig-2} shows the typical scission shapes obtained in the calculations for the three isotopes.
The most common scission configuration for $^{256}\text{Fm}$ corresponds to an asymmetric mass division
with a prolate light fragment and a slightly oblate heavy fragment.
This mode is identified as the standard fission mode introduced by Brosa~\cite{brosa1990}.
Fission of $^{258}\text{Fm}$ mainly leads to symmetric mass divisions with a very compact 
scission configuration with two spherical fragments.
This mode is identified as the super-short fission mode.
However, the standard fission mode seen in $^{256}\text{Fm}$ is also present in $^{258}\text{Fm}$ 
resulting in bimodal fission and relatively broad mass distribution in this nucleus.
In $^{260}\text{Fm}$ it is mainly the super-short fission mode that is appearing and the mass yield is therefore very narrow.

\begin{figure}[H]
\centering
\includegraphics[width=0.8\linewidth]{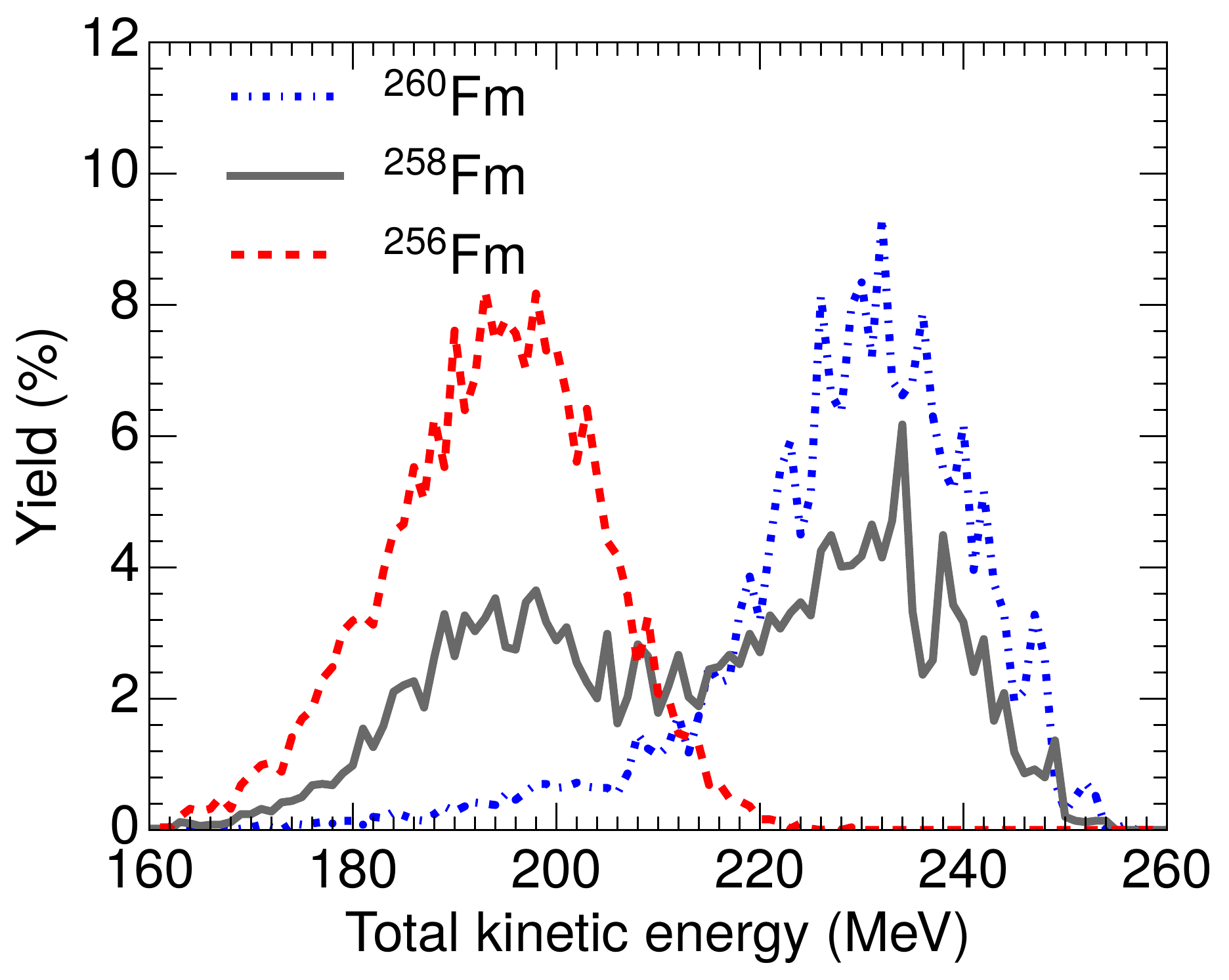}
\caption{Calculated TKE distributions for $^{256,258,260}\text{Fm}$.}
\label{fig-3}
\end{figure}

The calculated TKE distributions are shown in Fig.~\ref{fig-3}.
Since the TKE is largely determined by the strong Coulomb repulsion between the fragments,
the TKE value mainly depends on the elongation of the scission configuration.
The distribution for $^{256}\text{Fm}$ is well reproduced by a single Gaussian with an average TKE close to the measured value of 197.9 MeV
in spontaneous fission~\cite{hoffman1980}.
This TKE value is also close to the value expected by the Viola systematics for liquid-drop fission~\cite{viola1985}.
The distribution for $^{258}\text{Fm}$ on the other hand contains a mixture of the two modes, leading to a double-humped distribution.
Peaks near 200 and 235 MeV in the TKE distribution are also seen in measurements of spontaneous fission of $^{258}\text{Fm}$~\cite{hulet1986}.
The distribution for $^{260}\text{Fm}$ is peaked around 235 MeV due to the super-short fission mode, 
though a minor contribution from the standard mode is also visible.

Since the super-short fission mode is related to strong shell-effects, it has the most influence at low energies.
The high-TKE component in $^{258}\text{Fm}$ is therefore slightly less pronounced in the calculations, due to the excited nuclei, compared to
experimental data for spontaneous fission~\cite{hulet1986}.
When the excitation energy is increased, the shell-effects gradually disappear and the mass yields broaden, approaching what would be
obtained based on liquid-drop model potential energies.
The calculated TKE for both $^{258}\text{Fm}$ and $^{260}\text{Fm}$ then approaches the TKE value described by the standard asymmetric fission mode.

\vspace{-9mm}

\vspace{-5mm}

\begin{acknowledgement}
This work was supported by the Swedish Natural Science Research Council (S.{\AA}.) and 
the Knut and Alice Wallenberg Foundation Grant No. KAW 2015.0021 (M.A., B.G.C. and S.{\AA}.); 
J.R. was supported in part by the NNSA DNN R\&D of the U.S. Department of Energy.
\end{acknowledgement}

%
%
%

\vspace{-45mm}

\end{document}